\documentclass{article}
\usepackage{spconf,amsmath,graphicx}

\usepackage{gensymb}
\usepackage{tabularx,booktabs}
\newcolumntype{Y}{>{\centering\arraybackslash}X}
\usepackage[super]{nth}
\usepackage{hyperref}

\title{WHAT IS THE BEST DATA AUGMENTATION FOR 3D BRAIN TUMOR SEGMENTATION?}
 \name{Marco Domenico Cirillo$^{\: a,c}$, David Abramian$^{\: a,c}$, Anders Eklund$^{\: a,b,c}$}
 \address{$^a$Division of Medical Informatics, Department of Biomedical Engineering\\ $^b$Division of Statistics \& Machine Learning, Department of Computer and Information Science\\ $^c$Center for Medical Image Science and Visualization (CMIV) \\Link\"{o}ping University, Sweden}
\begin{document}
%
\maketitle
\begin{abstract}
Training segmentation networks requires large annotated datasets, which in medical imaging can be hard to obtain. Despite this fact, data augmentation has in our opinion not been fully explored for brain tumor segmentation. In this project we apply different types of data augmentation (flipping, rotation, scaling, brightness adjustment, elastic deformation) when training a standard 3D U-Net, and demonstrate that augmentation significantly improves the network's performance in many cases. Our conclusion is that brightness augmentation and elastic deformation work best, and that combinations of different augmentation techniques do not provide further improvement compared to only using one augmentation technique. Our code is available at \url{https://github.com/mdciri/3D-augmentation-techniques}.
\end{abstract}
\begin{keywords}
Data augmentation, 3D brain tumor segmentation, MRI, 3D U-Net, deep learning, artificial intelligence
\end{keywords}
\section{Introduction}
\label{sec:intro}
\vspace{-0.3cm}
Deep learning is increasingly being used in medical imaging, as it often provides better results compared to traditional analysis methods~\cite{litjens2017survey}. Compared to other computer vision tasks, it is in medical imaging more difficult to acquire a large number of training images due to ethics and data protection regulations (e.g. GDPR), and data augmentation is therefore even more important to increase the number of images for training and testing. Surprisingly, there are very few papers that investigate how important different types of augmentation (e.g. rotations, random flipping, scaling, elastic deformations) are for training convolutional neural networks (CNNs) for image classification or image segmentation, and this is especially true for 3D CNNs.
In \cite{menze} it is reported that the number of papers about brain tumor segmentation has been increasing exponentially in the last decade, and most of them utilizing artificial intelligence. Moreover, just considering the past Multimodal Brain tumor Segmentation (BraTS) Challenges from 2017 to 2020 the number of participants were 34, 64, 93, and 55 respectively. Many of these papers briefly mention that data augmentation was used, but do not mention details like the range of the random rotations or elastic deformations, or if the random rotations were drawn from a normal distribution or from a uniform distribution.

Nalepa et al. \cite{nalepa2019data} provide an overview of data augmentation for brain tumor segmentation, showing that flipping, rotation, scaling and pixel-wise (e.g. adding noise or changing the brightness) augmentation were most common, while translations and elastic deformations were less common. They also provide their own comparison of augmentation techniques, but it lacks details such how large the elastic deformations are. Furthermore, scaling was not tested, no statistical evaluation is done, and the used segmentation network seems to be 2D and not 3D. Shorten et al. \cite{shorten2019survey} provide a more general survey on data augmentation, but barely mention elastic deformations and brightness augmentation. A number of papers have proposed to learn the best augmentation \cite{cubuk2019autoaugment,xu2020automatic}, but a common drawback is a much longer training time. Here we therefore provide a comparison of different data augmentation techniques for brain tumor segmentation, so that new researchers in this field know what kind of augmentation to apply.

\section{Data}
\vspace{-0.3cm}
The MR images used for this project are from the BraTS Challenge 2020~\cite{menze,bakas1,bakas4}. The BraTS 2020 training dataset contains MR volumes of shape \(240\times240\times155\) from 369 patients, and for each patient four types of MR images were collected: native (T1), post-contrast T1-weighted (T1Gd), T2-weighted (T2), and T2 Fluid Attenuated Inversion Recovery (FLAIR). The BraTS 2020 validation dataset contains the same type of MR images from 125 patients, without the ground truth annotations. The images were acquired from 19 different institutions with different clinical protocols. The training set was segmented manually, by one to four raters, following the same annotation protocol, and their annotations were approved by experienced neuro-radiologists. Moreover, all data were co-registered to the same anatomical template, interpolated to the same resolution (1 mm\(^3\)) and skull-stripped. 

\section{Methods}
\vspace{-0.3cm}
In this project we use a standard 3D U-Net~\cite{ronneberger} architecture, one of the most used nowadays, which is trained with 4 MR images (T1, T1Gd, T2, FLAIR) to perform a 4 class segmentation: background, whole tumor (WT), tumor core (TC), and enhancing tumor (ET). Although more advanced segmentation networks have been proposed, a standard U-Net \cite{isensee} won the BraTS 2020 challenge. Our network is trained on sub-volumes of \(128\times128\times128\) voxels. The U-Net has 4 encoder and decoder steps: each step is made by two 3D convolutions: for the encoder a 3D convolution layer with stride 1 followed by another one with stride 2; whereas for the decoder a 3D Transpose convolution with stride 2, concatenated with the respective encoder step output, followed by a 3D convolution with stride 1. Each convolution has kernel size \(4\times4\times4\), He weights initialization \cite{he} and same padding. The first 3D convolution uses 64 filters which are doubled at each encoder step, viceversa for the decoder. Data pre- and post-processing were done exactly as in \cite{cirillo}. Adam \cite{kingma} is used as the network's optimizer with learning rate \(\lambda = 10^{-4}\), and the loss chosen is the generalized Dice loss \cite{sudre}. The augmentation techniques used for this projects are: 
\begin{itemize}
    \item \textit{Patch extraction}: from each original volume a sub-volume of shape \(128\times128\times128\) is extracted around its center. In this way each sub-volume mostly contains brain tissue and not the surrounding background.
    \vspace{-0.3cm}
    \item \textit{Flipping}: random flipping of one of the three different axes with 1/3 probability. 
    \vspace{-0.3cm}
    \item \textit{Rotation}: rotation applied to each axis with angles randomly chosen from a uniform distribution with range between 0\degree and 15\degree, 30\degree, 60\degree, or 90\degree.
    \vspace{-0.3cm}
    \item \textit{Scale}: scaling applied to each axis by a factor randomly chosen from a uniform distribution with range \(\pm10\)\% or \(\pm20\)\%. 
    \vspace{-0.3cm}
    \item \textit{Brightness}: power-law \(\gamma\) intensity transformation with its gain (\(g\)) and \(\gamma\) parameters chosen randomly between 0.8 - 1.2 from a uniform distribution. The intensity (\(I\)) is randomly changed according to the formula: \(I_{new} = g\cdot I^{\gamma}\).
    \vspace{-0.3cm}
    \item \textit{Elastic deformation}: elastic deformation with square deformation grid with displacements sampled from a normal distribution with standard deviation \(\sigma = 2\), 5, 8, or 10 voxels~\cite{ronneberger}, where the smoothing is done by a spline filter with order 3 in each dimension. 
\end{itemize}
Moreover, in order to report a robuster evaluation, 3-fold cross validation is applied to each model, and these 3 models are ensembled by averaging their softmax layer outputs.

Successively, the authors ranked all the augmentation approaches as in the BraTS 2020 challenge and handled the ties as in \cite{kendall1945treatment} to determine which of of these augmentation techniques and parameters yield the best performance on the validation set. There will be shown two rankings: one for all the different augmentation techniques and one including also their combinations.
Furthermore, the techniques with higher rank are also combined between each other with a probability of 0.5 for each patch. 

Each model is trained, with one or more augmentation techniques, over 200 epochs with early stopping after 25 epochs in case the validation loss does not decrease. The segmentation networks were trained with Nvidia Tesla V100 and Nvidia Quadro RTX 8000 graphics cards. In the end, all the trained models were evaluated on the 125 subjects of the BraTS 2020 challenge validation set. The metrics (Dice score and Hausdorff distance 95 percentile) are calculated by the CBICA Image Processing Portal (\url{https://ipp.cbica.upenn.edu}), while the rank scores were calculated by us.

\section{Results}
\begin{figure*}[h!]
    \centering
    \includegraphics[scale=0.345]{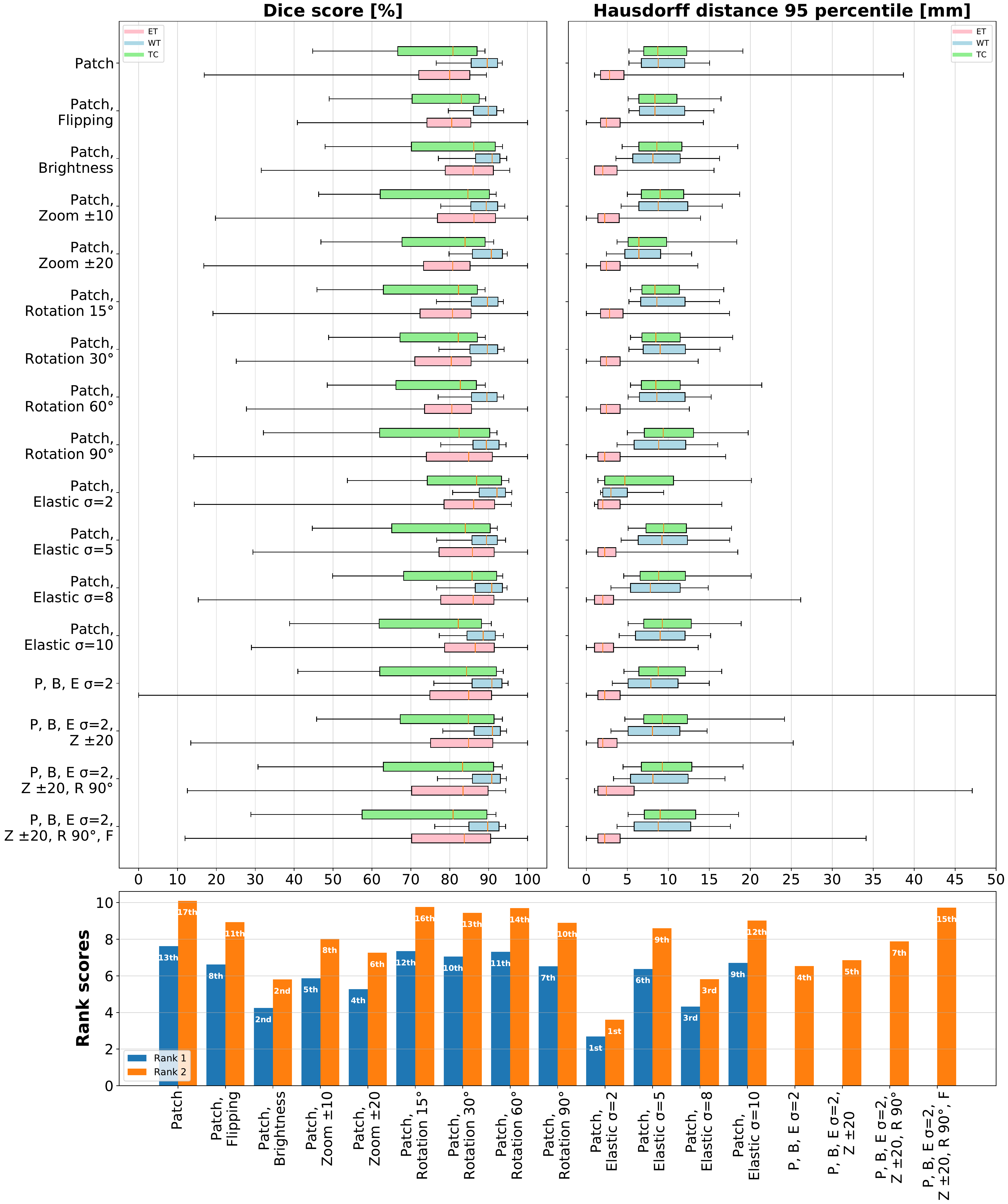}
    \caption{On top the Dice score and Hausdorff distance 95 percentile distributions over the different augmentation techniques applied during the 3D U-Net's training. Each box represents the IQR between the 25 and 75 percentile, the orange line inside the median, whereas the whiskers are set at 10 and 90 percentiles of the metrics for the ET, WT, and TC class highlighted in red, blue and green respectively. On the bottom the ranks scores calculated without (blue) and with (orange) considering multiple augmentation.}
    \label{fig:fig}
\end{figure*}

\vspace{-0.3cm}
First of all, we want to investigate whether data augmentation significantly increases the Dice score. Hence, we applied a non-parametric permutation test to the Dice scores from the 125 validation subjects. As the Dice scores are from the same 125 validation subjects in all cases, we used a paired t-test to test if the mean Dice difference is significantly different from zero. In non-parametric statistics, a paired t-test can be performed with a sign flipping test, where the sign of each pairwise difference is randomly flipped a large number of times, to obtain the null distribution~\cite{winkler2014permutation}. We used 100,000 sign flips per test and the p-values are given in Table~\ref{table:pvalues}. Brightness augmentation and elastic deformations with a $\sigma=2$ result in significantly higher Dice scores for all 3 tumor classes, scaling with $\pm$ 20\% significantly improves the Dice scores for 2 classes, while flipping and 90\degree rotation only significantly improve one tumor class.
\begin{table}[h!]
\scriptsize
\caption{Non-parametric p-values for comparing different types of data augmentation, obtained through a sign flipping test using the 125 validation subjects. ET = enhancing tumor, WT = whole tumor, TC = tumor core. The p-values have been multiplied with 36 (36 one sided tests) as Bonferroni correction for multiple comparisons. The bold numbers are those that are smaller than 0.05.}
\begin{center}
    \def\arraystretch{1.8}
    \begin{tabular}{|m{4cm}|m{1.0cm}m{1.0cm}m{1.0cm}|}
        \hline
         &   \multicolumn{3}{c|}{\textbf{p-value for Dice score}}  \\
        \textbf{Comparison}  & \textbf{ET} & \textbf{WT} & \textbf{TC}  \\
        \hline
        Flipping + PE $>$ PE & \textbf{0.0082} & 1.0 & 0.19116\\
        \hline
        Brightness + PE $>$ PE & \textbf{0.00036} & \textbf{0.00036} &  \textbf{0.00036}\\
        \hline
        Scale \(\pm10\)\% + PE $>$ PE & \textbf{0.00036} & 1.0 &  1.0\\
        \hline
        Scale \(\pm20\)\% + PE $>$ PE & 1.0 & \textbf{0.00036} &  \textbf{0.00396}\\
        \hline
        Rotation 0\degree - 15\degree + PE $>$ PE & 1.0 & 1.0 & 1.0\\
        \hline
        Rotation 0\degree - 30\degree + PE $>$ PE & 1.0 & 1.0 & 1.0\\
        \hline
        Rotation 0\degree - 60\degree + PE $>$ PE & 0.4676 & 0.2422 &  1.0\\
        \hline
        Rotation 0\degree - 90\degree + PE $>$ PE & \textbf{0.00036} & 1.0 & 1.0\\
        \hline
        Elastic deformation 2 + PE $>$ PE & \textbf{0.00036} & \textbf{0.00036} &  \textbf{0.00036}\\
        \hline
        Elastic deformation 5 + PE $>$ PE & \textbf{0.00036} & 1.0 & 0.44208\\
        \hline
        Elastic deformation 8 + PE $>$ PE & \textbf{0.00036} & \textbf{0.00071} & 0.13464\\
        \hline
        Elastic deformation 10 + PE $>$ PE & \textbf{0.00036} & 1.0 &  1.0\\
        \hline
    \end{tabular}
\end{center}
\label{table:pvalues}
\vspace{-0.6cm}
\end{table}
On top of Figure \ref{fig:fig} are illustrated the distributions of the Dice score and Hausdorff distance 95 percentile on the three different brain tumor classes over the 125 subjects in the validation set, for each tested augmentation approach. Each box represents the interquartile range (IQR) between the 25 and 75 percentile, the line inside the median, the whiskers are set at 10 and 90 percentiles of the metrics. On the other hand, on the bottom of Figure \ref{fig:fig} are represented the rank scores for the different augmentation techniques used, firstly without combining any technique, secondly including combinations of them with their best parameters. Looking at Figure \ref{fig:fig}, it is possible to say that brightness and elastic deformations with $\sigma=2$ are the two best augmentation techniques for this dataset, having the best data distributions and \nth{1} and \nth{2} position respectively. Moreover, the other most effective techniques are scaling with \(\pm20\)\%, rotation with random angle chosen between 0\degree - 90\degree, and flipping having \nth{4}, \nth{7}, and \nth{8} rank position respectively. Hence, four more trainings were done combining these techniques, and a second ranking was performed including them shown in Rank 2 in Figure \ref{fig:fig}. Brightness and elastic deformations with $\sigma=2$ kept their rank position, whereas the combination of them and the one including also scaling with \(\pm20\)\% take the \nth{4} and \nth{5} position respectively.
Note that the CBICA systems reports 0\% and 373mm as Dice score and Hausdorff 95 percentile when the network detects a class (usually the ET one) that is not present, or viceversa. In the case of the combination of elastic deformation with $\sigma=2$ and brightness, the ET whiskers go till 0\% and 373mm.

\section{Discussion}
\vspace{-0.3cm}
The aim of this paper is not to propose a state-of-the-art image segmentation method with high performance as in \cite{cirillo,isensee}, but to investigate how different augmentation techniques affect a CNN's learning. Our results show that data augmentation significantly improves the performance on the validation data in many cases compared to only using patch extraction as baseline technique (see Table~\ref{table:pvalues}). A possible explanation why data augmentation has not been fully explored for brain tumor segmentation is that the BraTS training set is rather large (369 subjects for the 2020 version), and several papers suggest that data augmentation would not help much~\cite{urban2014multi,lyksborg2015ensemble,havaei2017brain}. Since all subjects in BraTS have been registered to a common space, augmentation is important to show the network brains from different angles, while augmentation may not be as important for a dataset where the brains have not been registered.

In Figure \ref{fig:fig} reports also the rank scores for our tested approaches. It is clear that augmentation improves the training performance because patch augmentation, which is our baseline, has the lowest rank in both cases (\nth{13} and \nth{17}). Augmentation techniques have to be chosen smartly in order to create new training images that still represent and/or look like the original ones. For this particular dataset:
\begin{itemize}
    \item elastic deformation generates realistic tumors and achieves its best scores with \(\sigma=2\), which is also the technique with the best rank position. Additionally, even using a value of \(\sigma=5\) or 8, the scores are still suitable, achieving the \nth{6} and \nth{3} rank position respectively, whereas \(\sigma=10\) it is too high and it deforms the brains excessively;
    \vspace{-0.3cm}
    \item \(\gamma\) correction (brightness) augmentation is the second best form of augmentation here, which is probably explained by the fact that the BraTS 2020 dataset contains data from 19 different sites. Moreover, according to Nalepa \textit{et al.} \cite{nalepa2019data}, only Isensee \textit{et al.}~\cite{isensee} used gamma correction in their comparison of the BraTS 2018 participants;
    \vspace{-0.3cm}
    \item random scaling and rotation create bigger and/or smaller tumors that are distributed everywhere in the training volumes' space. This is probably the reason why greater scaling and rotation ranges result in better scores;
    \vspace{-0.3cm}
    \item flipping is very easy to implement technique, but it does not increase significantly the network's performance.
\end{itemize}
Moreover, Figure \ref{fig:fig} shows that the best combination is the one with elastic deformation with $\sigma=2$ and brightness, followed by the one with scaling $\pm20$, but it is interesting to notice that combining different augmentation techniques does not improve the network's performance. Indeed, the models trained with only elastic deformation or brightness augmentation are higher ranked than the ones trained on multiple augmentation techniques (see Rank 2). This may be due to the fact that each augmentation technique is applied with a probability 0.5, so, if more augmentation techniques are applied, fewer original images are shown to the network during the training. Indeed, the percentage of original images used during the training is 50\%,  25\%, 12.5\%, 6.25\%, and 3.125\% combining 1, 2, 3, 4 and 5 different techniques respectively. Combining different types of augmentation can potentially be done differently, to always guarantee that for example 20\% of the shown patches are always the original ones.

To summarize, the authors recommend everyone who is going to use this type of dataset in the future to apply, at least, elastic deformation and brightness adjustment as augmentation techniques, and in case they would like to use more augmentation techniques, to be sure that the network considers a suitable percentage of original images during its training.


\section{Acknowledgments}
\vspace{-0.3cm}
This study was supported by LiU Cancer, VINNOVA Analytic Imaging Diagnostics Arena (AIDA), and the ITEA3 / VINNOVA funded project Intelligence based iMprovement of Personalized treatment And Clinical workflow supporT (IMPACT). 
Anders Eklund has previously received hardware from Nvidia, otherwise the authors have no conflicts of interest to declare.


\bibliographystyle{IEEEbib}
\bibliography{strings}

\begin{thebibliography}{10}

\bibitem{litjens2017survey}
Geert Litjens, Thijs Kooi, Babak~Ehteshami Bejnordi, Arnaud Arindra~Adiyoso
  Setio, Francesco Ciompi, Mohsen Ghafoorian, Jeroen~Awm Van Der~Laak, Bram
  Van~Ginneken, and Clara~I S{\'a}nchez,
\newblock ``A survey on deep learning in medical image analysis,''
\newblock {\em Medical image analysis}, vol. 42, pp. 60--88, 2017.

\bibitem{menze}
Bjoern~H Menze, Andras Jakab, Stefan Bauer, Jayashree Kalpathy-Cramer, Keyvan
  Farahani, Justin Kirby, Yuliya Burren, Nicole Porz, Johannes Slotboom, Roland
  Wiest, et~al.,
\newblock ``The multimodal brain tumor image segmentation benchmark
  ({BRATS}),''
\newblock {\em IEEE transactions on medical imaging}, vol. 34, no. 10, pp.
  1993--2024, 2014.

\bibitem{nalepa2019data}
Jakub Nalepa, Michal Marcinkiewicz, and Michal Kawulok,
\newblock ``Data augmentation for brain-tumor segmentation: A review,''
\newblock {\em Frontiers in Computational Neuroscience}, vol. 13, 2019.

\bibitem{shorten2019survey}
Connor Shorten and Taghi~M Khoshgoftaar,
\newblock ``A survey on image data augmentation for deep learning,''
\newblock {\em Journal of Big Data}, vol. 6, no. 1, pp. 60, 2019.

\bibitem{cubuk2019autoaugment}
Ekin~D Cubuk, Barret Zoph, Dandelion Mane, Vijay Vasudevan, and Quoc~V Le,
\newblock ``Autoaugment: Learning augmentation strategies from data,''
\newblock in {\em Proceedings of the IEEE conference on computer vision and
  pattern recognition}, 2019, pp. 113--123.

\bibitem{xu2020automatic}
Ju~Xu, Mengzhang Li, and Zhanxing Zhu,
\newblock ``Automatic data augmentation for {3D} medical image segmentation,''
\newblock in {\em International Conference on Medical Image Computing and
  Computer-Assisted Intervention}. Springer, 2020, pp. 378--387.

\bibitem{bakas1}
Spyridon Bakas, Hamed Akbari, Aristeidis Sotiras, Michel Bilello, Martin
  Rozycki, Justin~S Kirby, John~B Freymann, Keyvan Farahani, and Christos
  Davatzikos,
\newblock ``Advancing the cancer genome atlas glioma {MRI} collections with
  expert segmentation labels and radiomic features,''
\newblock {\em Scientific data}, vol. 4, pp. 170117, 2017.

\bibitem{bakas4}
Spyridon Bakas, Hamed Akbari, Aristeidis Sotiras, Michel Bilello, Martin
  Rozycki, Justin Kirby, John Freymann, Keyvan Farahani, and Christos
  Davatzikos,
\newblock ``Segmentation labels and radiomic features for the pre-operative
  scans of the {TCGA-LGG} collection,''
\newblock {\em The Cancer Imaging Archive}, 2017.

\bibitem{ronneberger}
Olaf Ronneberger, Philipp Fischer, and Thomas Brox,
\newblock ``U-net: Convolutional networks for biomedical image segmentation,''
\newblock in {\em International Conference on Medical image computing and
  computer-assisted intervention}. Springer, 2015, pp. 234--241.

\bibitem{isensee}
Fabian Isensee, Philipp Kickingereder, Wolfgang Wick, Martin Bendszus, and
  Klaus~H Maier-Hein,
\newblock ``No new-net,''
\newblock in {\em International MICCAI Brainlesion Workshop}. Springer, 2018,
  pp. 234--244.

\bibitem{he}
Kaiming He, Xiangyu Zhang, Shaoqing Ren, and Jian Sun,
\newblock ``Delving deep into rectifiers: Surpassing human-level performance on
  imagenet classification,''
\newblock in {\em Proceedings of the IEEE international conference on computer
  vision}, 2015, pp. 1026--1034.

\bibitem{cirillo}
Marco~Domenico Cirillo, David Abramian, and Anders Eklund,
\newblock ``Vox2{V}ox: 3{D-GAN} for {B}rain {T}umour {S}egmentation,''
\newblock in {\em International MICCAI Brainlesion Workshop}. Springer, 2020.

\bibitem{kingma}
Diederik~P Kingma and Jimmy Ba,
\newblock ``Adam: A method for stochastic optimization,''
\newblock {\em arXiv preprint arXiv:1412.6980}, 2014.

\bibitem{sudre}
Carole~H Sudre, Wenqi Li, Tom Vercauteren, Sebastien Ourselin, and M~Jorge
  Cardoso,
\newblock ``Generalised {D}ice overlap as a deep learning loss function for
  highly unbalanced segmentations,''
\newblock in {\em Deep learning in medical image analysis and multimodal
  learning for clinical decision support}, pp. 240--248. Springer, 2017.

\bibitem{kendall1945treatment}
Maurice~G Kendall,
\newblock ``The treatment of ties in ranking problems,''
\newblock {\em Biometrika}, pp. 239--251, 1945.

\bibitem{winkler2014permutation}
Anderson~M Winkler, Gerard~R Ridgway, Matthew~A Webster, Stephen~M Smith, and
  Thomas~E Nichols,
\newblock ``Permutation inference for the general linear model,''
\newblock {\em Neuroimage}, vol. 92, pp. 381--397, 2014.

\bibitem{urban2014multi}
Gregor Urban, M~Bendszus, F~Hamprecht, and J~Kleesiek,
\newblock ``Multi-modal brain tumor segmentation using deep convolutional
  neural networks,''
\newblock {\em MICCAI BraTS (brain tumor segmentation) challenge. Proceedings,
  winning contribution}, pp. 31--35, 2014.

\bibitem{lyksborg2015ensemble}
Mark Lyksborg, Oula Puonti, Mikael Agn, and Rasmus Larsen,
\newblock ``An ensemble of {2D} convolutional neural networks for tumor
  segmentation,''
\newblock in {\em Scandinavian Conference on Image Analysis}. Springer, 2015,
  pp. 201--211.

\bibitem{havaei2017brain}
Mohammad Havaei, Axel Davy, David Warde-Farley, Antoine Biard, Aaron Courville,
  Yoshua Bengio, Chris Pal, Pierre-Marc Jodoin, and Hugo Larochelle,
\newblock ``Brain tumor segmentation with deep neural networks,''
\newblock {\em Medical image analysis}, vol. 35, pp. 18--31, 2017.

\end{thebibliography}

\end{document}